# New radio lobes at parsec scale from the East-West protostellar jet RAFGL2591

A. G. Cheriyan[1]⋆, S. Vig[1] and Sreelekshmi Mohan[1]

[1] *Indian Institute of Space science and Technology, Thiruvananthapuram 695547, India*

**ABSTRACT**

RAFGL2591 is a massive star-forming complex in the Cygnus-X region comprising of a cluster of embedded protostars and young stellar objects located at a distance of 3.33 kpc. We investigate low-frequency radio emission from the protostellar jet associated with RAFGL2591 using the Giant Metrewave Radio Telescope (GMRT) at 325, 610 and 1280 MHz. For the first time, we have detected radio jet lobes in the E-W direction, labelled as GMRT-1 and GMRT-2. While GMRT-1 displays a flat radio spectral index of $\alpha = -0.10$, GMRT-2 shows a steeply negative value $\alpha = -0.62$ suggestive of non-thermal emission. $H_2$ emission maps show the presence of numerous knots, arcs and extended emission towards the East-West jet, excited by the protostar VLA 3. In addition, we report a few $H_2$ knots in the North-East and South-West for the first time. The radio lobes (GMRT-1, GMRT-2) and $H_2$ emission towards this region are understood in the context of the prominent East-West jet as well as its lesser-known sibling jet in the North-East and South-West direction. To model the radio emission from the lobes, we have employed a numerical model including both thermal and non-thermal emission and found number densities towards these lobes in the range $100 - 1000$ cm$^{-3}$. The misalignment of the East-West jet lobes exhibits a reflection symmetry with a bending of $\sim 20°$. We attempt to understand this misalignment through precession caused by a binary partner and/or a supersonic side wind from source(s) in the vicinity.

## 1 INTRODUCTION

Jets and outflows are omnipresent in protostars, and they assist in the removal of excess angular momentum from the protostellar systems. In addition, it is believed that the cavities that they produce aid in venting the high radiation pressure from massive protostars (Anglada et al. 2018). Jets and the gas that they entrain as outflows have been employed as crucial kinematic tools in locating young protostars as they are embedded within massive clouds of gas and dust (Arce et al. 2007). It is believed that jets and outflows affect the parental molecular cloud, as they play a role in inducing turbulence within the cloud, which could plausibly trigger or hinder the formation of stars in the vicinity (Li & Nakamura 2006; Nakamura & Li 2007; Shimajiri et al. 2008).

The presence of jets and outflows represent a notable signpost in the protostellar evolution that has been utilised for categorising evolutionary stages of young stellar objects (Beuther & Shepherd 2005; Schulz 2012). The launching mechanism of jets is still under debate; however, what is known with certainty is that magnetic fields play an important role in the launch and collimation, where the toroidal component of the helical magnetic field is responsible for the confining pressure leading to collimation of jets (Livio 1997; Meier et al. 2001). This has been substantiated by several magneto-hydrodynamic models (Ray & Ferreira 2021; Shang et al. 2023a,b).

Jets have been observed across a multitude of wavebands from radio to X-rays (Ray & Ferreira 2021; Bally 2016). In this paper, we focus on the radio emission associated with jets. The radio continuum emission from jets can be categorised as thermal (free-free) or non-thermal (synchrotron) based on the spectral index (Anglada 1995). The spectral index $\alpha$ represents the power-law index of the change of flux density of emission, $S_\nu$, with respect to frequency $\nu$, and can be written as $S_\nu \propto \nu^\alpha$. Regions where the spectral index is greater than $\alpha \gtrsim -0.1$, are presumed to be dominated by thermal emission, (Olnon 1975), and when the spectral index is negative with $\alpha \lesssim -0.5$, then the dominant emission is due to synchrotron process (Terquem et al. 1999). The latter has been confirmed through the presence of polarized radio emission from jets (Carrasco-González et al. 2010; Cécere et al. 2016). If both the mechanisms are at play, then the spectral indices are envisaged to be flat as the power-law slopes have opposite signs with respect to the frequency for both processes (Rodríguez-Kamenetzky et al. 2016).

The non-thermal emission is believed to arise from the shocked regions of jets, which are created when they interact with the ambient medium (Rodríguez-Kamenetzky et al. 2016; Vig et al. 2018; Obonyo et al. 2019). This has been observed towards multiple massive young stellar objects (Vig et al. 2018; Masqué et al. 2019). The particles in these shocked regions get accelerated to relativistic velocities and radiate synchrotron emission in the presence of a magnetic field generated by changes in protostellar plasma and amplified through the dynamo effect (Latif & Schleicher 2016). Due to the negative spectral index, the non-thermal emission can best be observed at low radio frequencies. This non-thermal emission can be used to infer the properties of the ionised gas associated with the jet such as ionization fraction, and mass-loss rate, as well as of the ambient medium in the vicinity.

⋆ E-mail: amalgeorgecheriyan@gmail.com

In this work, we investigate radio emission from the protostellar jet associated with RAFGL2591. This high-mass star-forming region is located in the Cygnus-X region of the sky at a distance of 3.33 kpc (Rygl et al. 2012). The region hosts a cluster of protostars and massive young stellar objects that give rise to multiple HII regions and outflows (Trinidad et al. 2003; Gieser et al. 2019). On large scales, a molecular outflow has been observed in CO, oriented approximately in the East-West direction (Lada et al. 1984; Gieser et al. 2019), which is believed to be excited by the radio source VLA-3 that is associated with the bright near-infrared source NIRS1 (Tamura & Yamashita 1992; Johnston et al. 2013). In the near-infrared K-band, several loops are observed, which are aligned along the blue CO outflow lobe towards the west of RAFGL2591, that are plausibly outflow cavities formed as a result of episodic mass ejections that have entrained the surrounding material (Arce & Sargent 2004). A three-colour composite image of the RAFGL2591 region with images of J (1.2 $\mu$m), H (1.6 $\mu$m) and K (2.1 $\mu$m) bands using the United Kingdom Infrared Telescope (UKIRT) which is shown in Fig. 1, with CO emission overlaid as contours. Shocked $H_2$ emission knots and Herbig Haro (HH) objects in [SII] line have also been observed along the outflow direction (Poetzel et al. 1992; Tamura & Yamashita 1992). High-resolution sub-millimetre continuum observations have demonstrated the presence of three fragments A, B and C, plausibly the outcome of disk-fragmentation, with subsolar masses within the inner 1000 au of VLA 3 (Suri et al. 2021).

In radio, several VLA continuum compact sources have been detected in the region, which are associated with HII regions, hot cores and maser emission (Johnston et al. 2013; Sanna et al. 2012). The VLA sources are designated as VLA 1-5, of which VLA 1, 2, 4 and 5 are H-II regions. VLA 4 and 5 are associated with infrared sources (Trinidad et al. 2003; Johnston et al. 2013) and VLA 3 is a massive hot core of mass 40 $M_\odot$ at 1.3 millimetre wavelength with 0.″4 resolution (Beuther et al. 2018). The 3.6 cm VLA observations have confirmed the morphology of VLA 3 to be a combination of a compact core and collimated jet (Johnston et al. 2013). The CO outflow seen at larger scales is almost parallel to the jet, thus confirming VLA 3 as the powering source for the outflow (Hasegawa & Mitchell 1995; Johnston et al. 2013).

Our motivation in the current work is to observe the protostellar jet at low radio frequencies in our quest for non-thermal emission in the region. In particular, we examine the low-frequency radio (325-1280 MHz) characteristics of the jet-driven by VLA 3. In addition, we search for shocked gas associated with the jet in the near-infrared tracer of $H_2$. The structure of the paper is organized as follows. The details of observations and data reduction are given in Sect. 2. We present our results in Sect. 3, which is discussed in Sect. 4. A succinct overview of our conclusions is provided in Sect. 5.

## 2 OBSERVATIONS AND ARCHIVAL DATA

### 2.1 Radio continuum observations using GMRT

Low radio frequency observations of the star-forming region RAFGL2591 have been carried out in three frequency bands: Band

**Table 1.** Details of the GMRT observations of RAFGL2591.

| Frequency (MHz) | 325 | 610 | 1280 |
|---|---|---|---|
| Observation date | 01 May 2022 | 02 August 2021 | 01 May 2022 |
| On source time (hrs) | 7 | 7.5 | 6 |
| Bandwidth (MHz) | 200 | 400 | 400 |
| Primary beam (′) | 86.0 | 45.8 | 21.5 |
| Synthesized beam (″)$^2$ | 7.6 × 6.2 | 7.0 × 5.2 | 1.8 × 1.5 |
| Position angle (°) | 58.58 | 29.73 | 42.16 |
| Noise ($\mu$Jy/beam) | 80 | 60 | 40 |

3 (250-500 MHz), Band 4 ( 550-850 MHz) and Band 5 (1050-1450 MHz) using the Giant Metrewave Radio Telescope (GMRT) located near Pune, India. The GMRT comprises of 30 antennas, each with a diameter of 45 m, arranged in a Y-shaped configuration. Twelve antennas are placed randomly within a central region measuring $\sim 1 \times 1$ square kilometre, and 18 antennas are located along three arms, each about 14 km long (Swarup et al. 1991). The shortest and longest baselines are 105 meters and 25 kilometres, respectively, simultaneously enabling the mapping of both large- and small-scale structures. At 325, 610, and 1280 MHz, the angular sizes of the largest observable structures are 32′, 17′, and 7′, respectively. Our initial observations were carried out at 610 MHz in August 2021, followed by 325 and 1280 MHz in May 2022. The details of the observations are listed in Table 1. With the upgraded wide-band receivers of GMRT, our observations included a bandwidth of 100 MHz at 325 MHz, 200 MHz at 610 MHz and 280 MHz at 1280 MHz. The flux calibrators 3C286 and 3C48 were observed at the beginning and end of observations to establish the flux density scale. To keep track of the phase and amplitude fluctuations associated with the target source, the phase calibrator 2015+371 was observed for five minutes after every scan of 40 min of RAFGL2591 observation.

The data reduction was carried out using the NRAO Astronomical Image Processing System (AIPS). The data in all frequency bands were partitioned into datasets, each of bandwidth 32 MHz, to reduce the effect of bandwidth smearing during the analysis. These datasets were carefully checked for corrupted data due to radio frequency interference, non-working antennas, etc., and were flagged using various tasks in AIPS. Following this, the datasets were calibrated, and the data for the RAFGL2591 was segregated. These were then cleaned and deconvolved using the task `IMAGR`. Several iterations of self-calibration were applied to minimize phase errors. The primary beam was corrected using the task `PBCOR` with coefficients obtained from GMRT primary beam shapes. Finally, all the cleaned images were combined to obtain the final radio continuum map. The final synthesized beams achieved are 7″.6 × 6″.2, 7″.0 × 5″.2, and 1″.8 × 1″.5 at 325, 610 and 1280 MHz frequencies, respectively (also tabulated in Table 1). The noise in the final maps is 80, 60 and 40 $\mu$Jy/beam at 325, 610 and 1280 MHz, respectively.

### 2.2 Archival data: UWISH2 Survey

We have utilised the 2.12 $\mu$m continuum-subtracted images from UKIRT Widefield Infrared Survey for $H_2$ (UWISH2) (Froebrich et al. 2011) for comparison with emission from the radio continuum images. The images are obtained using the Wide Field Camera (WF-CAM) on the United Kingdom Infrared Telescope (UKIRT), Mauna Kea, Hawaii. The point spread function is ~0″.6 at 2 $\mu$m, and the pixel scale in images measures 0″.4. The exposure time was 720 s for the source of our interest with a field-of-view of ~ 0.75 square degrees.

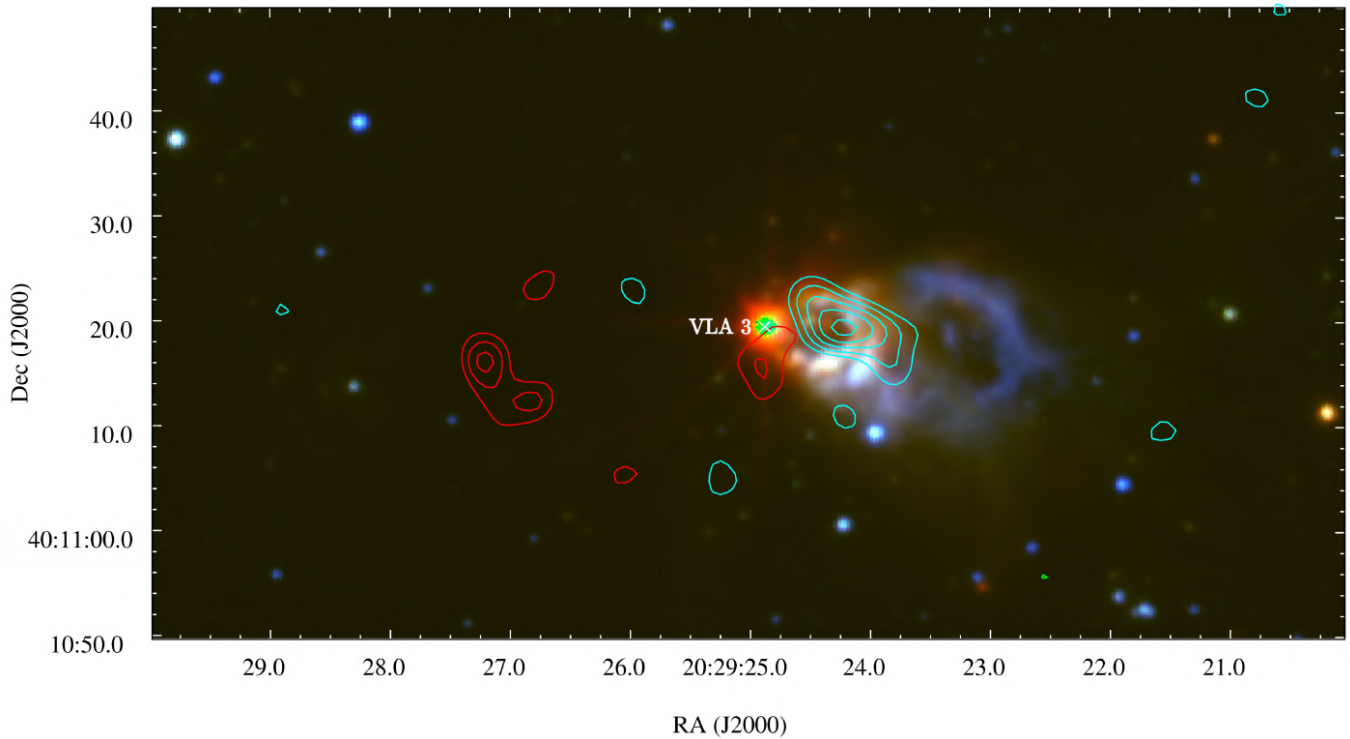

**Figure 1.** The UKIRT Infrared Deep Sky Survey (UKIDSS) three-colour composite image of RAFGL2591 region with J (blue), H (green), and K (red) band images. The contours represent the $C^{18}O$ emission integrated over $-4.3$ to $-3.3$ km/s (red) and $-8.0$ to $-7.0$ km/s (blue) taken from Johnston et al. (2013).

## 3 RESULTS

### 3.1 Ionised gas emission

Figure 2 displays the ionized continuum emission from the RAFGL2591 region at low radio frequencies at Band 3, Band 4, and Band 5 observed using GMRT, hereafter referred to as 325, 610 and 1280 MHz, respectively, for ease of frequency identification. At the highest resolution 1280 MHz band, we note the presence of the clustered ionised regions, designated VLA 1 to VLA 5 (Campbell 1984; Trinidad et al. 2003; Johnston et al. 2013). The flux densities of the VLA sources have been estimated by selecting elliptical apertures. The size of the elliptical apertures is determined by visual inspection. The errors in flux densities are estimated using the expression $\sqrt{\left(2\sigma\sqrt{\theta_{src}/\theta_{bm}}\right)^2 + (2\sigma')^2}$. Here, $\theta_{bm}$ stands for the size of the beam and $\theta_{src}$ represents the source size, which is calculated as the geometric mean of the major and minor axes of the aperture (Sánchez-Monge et al. 2013). The GMRT uncertainty in the flux calibration is taken as 5 percent (Lal & Rao 2007). The peak and integrated flux densities are listed in Table 2. The brightest emission is observed towards VLA 1. We note that the shapes of VLA 1, VLA 2, VLA 4 and VLA 5 appear spherical, while VLA 3 is elongated in the East-West direction. At 610 MHz, we clearly discern VLA 4 and VLA 5, but the other three sources are not well-resolved, although we note an extension corresponding to VLA 3. In the lower resolution 325 MHz image, apart from VLA 1, the other sources are not resolved, although we observe extensions towards VLA 4 and VLA 5.

In addition to the cluster of ionised regions at the centre, we perceive bright emission at a distance of $\sim 40''$ from VLA 3 towards the East and West directions in the GMRT maps. We label them as GMRT-1 and GMRT-2, respectively. At 1280 MHz, we note substructures in these objects. GMRT-1 shows the presence of three compact regions of ionised gas emission in addition to extended cometary-shaped emission around it, with the head pointed towards the east. Towards GMRT-2, we observe bright, compact emission and associated nebulous tail-like emission. At 610 and 325 MHz, the emission from GMRT-1 and GMRT-2 are clearly seen. Based on their close association with $H_2$ knots, we believe that GMRT-1 and GMRT-2 are associated with bow-shocks from the protostellar jet associated with VLA 3 (see Sect. 3.2). This is the first time that radio emission from lobes is detected at distances of $\sim 0.6$ pc from VLA 3.

RAFGL2591 has been imaged earlier at higher frequencies using the Very Large Array (VLA). We have compared the emission from GMRT with the VLA image at 8.4 GHz, having a resolution of $\sim 0.4''$ (Johnston et al. 2013); their images show only the central region. In order to search for associated emission at 8.4 GHz towards GMRT-1 and GMRT-2, we extracted and analysed their VLA data (Proposal ID: AJ337) using AIPS. The 8.4 GHz image is shown in Fig. 3. We find two compact sources of emission towards GMRT 1, at the level of $15\sigma$ and $6\sigma$, where $\sigma = 25$ $\mu$Jy/beam is the rms noise. Towards GMRT-2, we detect a compact source at the level of $6\sigma$. Thus, we note that the ionised outflow lobes detected with GMRT are also observed at a higher frequency of 8.4 GHz.

### 3.2 $H_2$ emission in the region: Association with GMRT-1 and GMRT-2

Regions of shocked gas associated with the interaction of jets and outflows with the ambient medium can be traced using the $H_2$ emission lines in the near-infrared (Shull & Beckwith 1982; Bally & Lane 1991). Taking advantage of the high resolution achieved in near-infrared images, it is possible to probe detailed substructures in regions of shock-excited gas using the $H_2$ $v = 1 - 0$ S(1) line. The higher photon energy of the ro-vibrational transition ensures that the energy levels from which the observed lines originate are only activated in hot gas ($T > 10^3$ K) found in shocks or UV-irradiated regions (Bally & Lane 1991). We employ the continuum subtracted $H_2$ $v = 1 - 0$ S(1) image from the UWISH2 towards RAFGL2591, shown in Fig. 4, to analyse the regions of $H_2$ emission. We note the presence of several arcs, filamentary structures and knots in regions away from the exciting source(s), marking them as locations of shock-excitation. In an earlier work, Tamura & Yamashita (1992) had detected emission knots towards this region and labelled them as A-F. We carry out a comparison of these knots with the $H_2$ structures seen in the UWISH2 image.

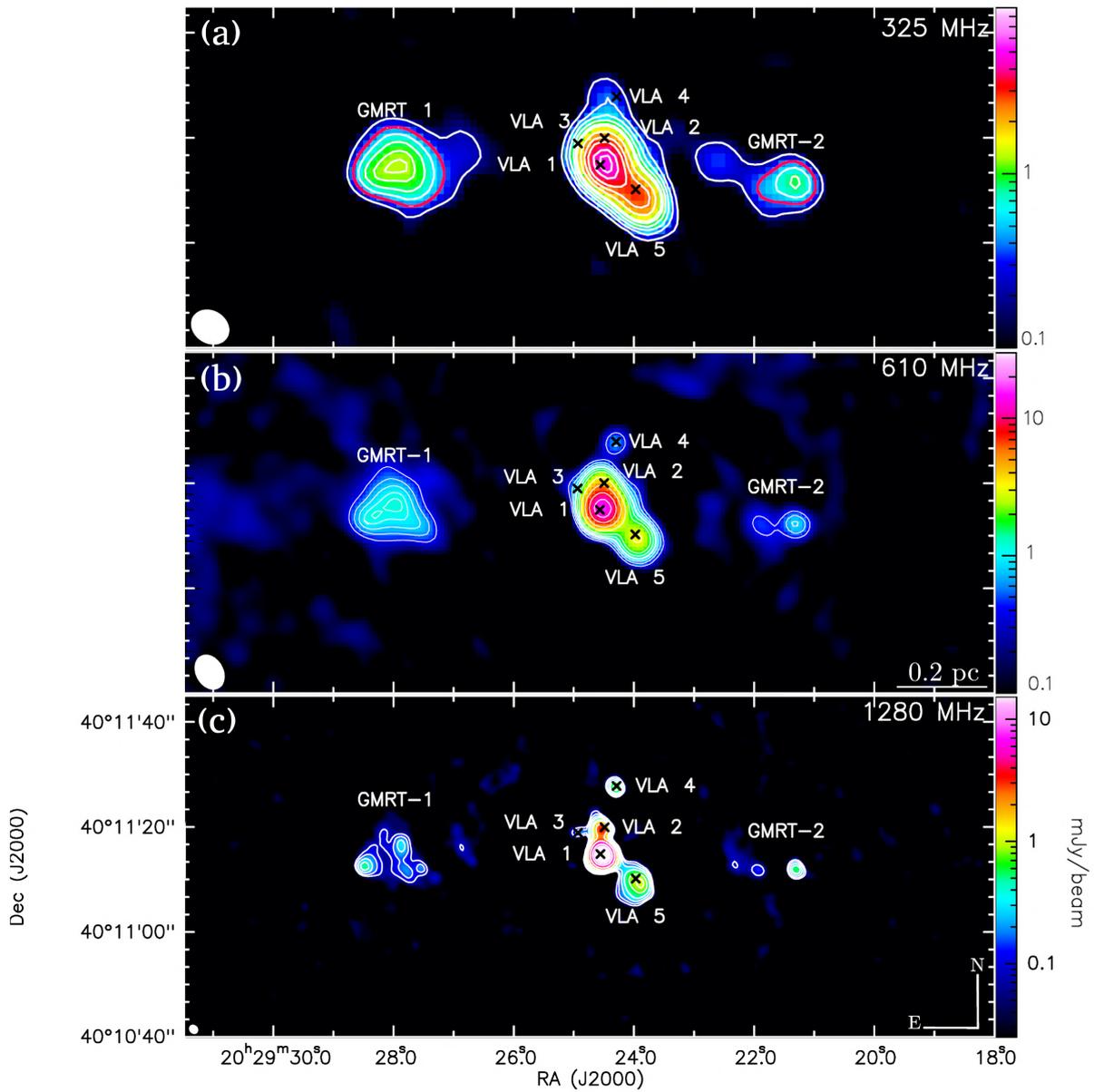

**Figure 2.** Radio maps of RAFGL2591 region at 325, 610 and 1280 MHz obtained from the GMRT. The contour levels are at the following intensity levels: 325 MHz - 0.2, 0.4, 0.6, 0.85, 1.1, 1.25, 1.8, 2.4, 3.0, 4.0, and 4.7 mJy/beam, where beam: $7.''0 \times 5.''8$, 610 MHz - 0.4, 0.6, 0.8, 1.0, 1.1, 1.8, 3.0, 5.0, 8.0, 12, and 14 mJy/beam, where beam: $4.''4 \times 4.''1$ and 1280 MHz - 0.09, 0.15, 0.25, 0.40, 0.70, 1.0, 2.0, 4.0, 8.0, 12, and 16 mJy/beam, where beam: $1.''8 \times 1.''1$. The apertures used for estimating flux densities are overlaid in red on the 325 MHz image.

**Table 2.** Details of VLA sources in RAFGL2591 at 1280 and 610 MHz, including the flux densities.

| Source | Frequency (MHz) | $\alpha_{J2000}$ (hms) | $\delta_{J2000}$ (° ′ ″) | $\theta_{src}$ (″×″) | Peak flux density (mJy beam$^{-1}$) | Int. flux density (mJy) |
|---|---|---|---|---|---|---|
| VLA 1 | 1280 | 20:29:24.55 | +40:11:14.59 | 3.7 × 3.7 | 26.1 ± 0.2 | 81.1 ± 7.9 |
| VLA 2 | 1280 | 20:29:24.59 | +40:11:20.31 | 2.1 × 2.0 | 3.6 ± 0.3 | 8.2 ± 0.8 |
| VLA 3 | 1280 | 20:29:24.97 | +40:11:19.27 | 2.3 × 1.3 | 0.70 ± 0.06 | 0.80 ± 0.07 |
| VLA 1-3 | 610 | 20:29:24.51 | +40:11:16.70 | 7.0 × 7.0 | 2.1 ± 0.1 | 24.3 ± 2.1 |
| VLA 4 | 1280 | 20:29:24.33 | +40:11:27.90 | 1.7 × 2.0 | 0.90 ± 0.08 | 1.4 ± 0.1 |
|  | 610 | 20:29:24.33 | +40:11:28.00 | 3.3 × 3.4 | 0.1 ± 0.03 | 0.50 ± 0.04 |
| VLA 5 | 1280 | 20:29:24.00 | +40:11:09.26 | 3.7 × 4.0 | 1.1 ± 0.1 | 8.1 ± 0.7 |
|  | 610 | 20:29:23.90 | +40:11:08.20 | 4.8 × 4.8 | 4.1 ± 0.4 | 6.2 ± 0.6 |

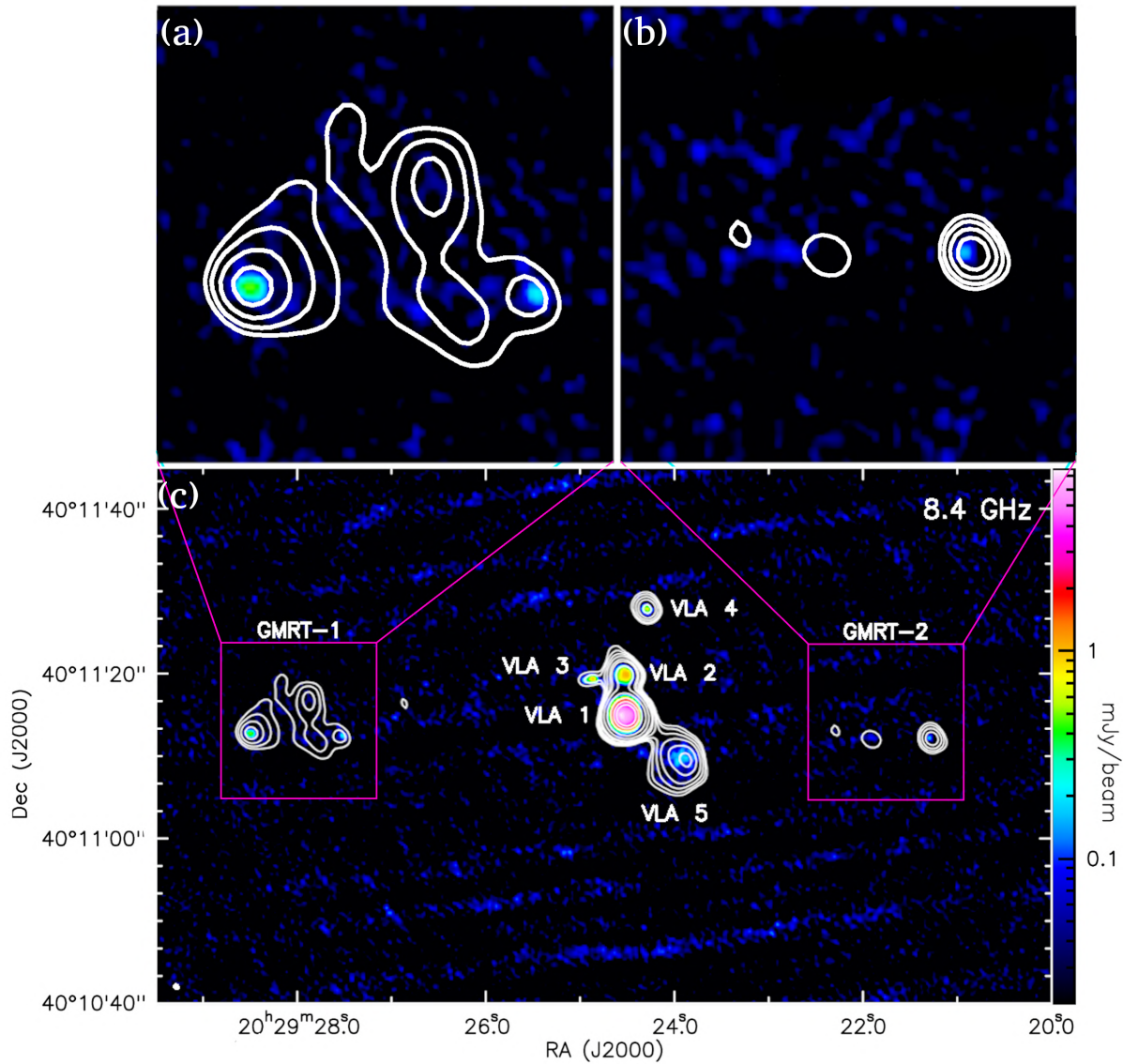

**Figure 3.** [Bottom panel]: VLA radio image of RAFGL2591 region at 8.4 GHz overlaid with contours of 1280 MHz emission from GMRT, contour levels same as shown in Fig. 2. The enlarged insets enclosing GMRT-1 and GMRT-2 are displayed on top.

We broadly demarcate three regions associated with $H_2$ emission for convenience: Region 1 corresponds to the central region, while Regions 2 and 3 cover GMRT-1 and GMRT-2, respectively. These are shown in Fig. 4 (a), (b) and (c), respectively. Region 1 includes the central embedded cluster of ionised regions, i.e. VLA sources. We note that VLA 3 is saturated in the image with the presence of a few $H_2$ knots in the vicinity. In addition to knots B and F, we identify three additional knots here. There are two bright knots towards the North-East of VLA 3, and we label them as H, and I. These are located at distances of 24″ (0.38 pc) and 26″ (0.41 pc) from VLA 3, respectively. We also identify the presence of a knot to the west of knot F; this is labelled as G and is located at a distance of 15″ (0.24 pc) from VLA 3. No excess $H_2$ emission is detected towards the location of knot B in the UWISH2 image. However, there is a bright arc towards the South-East, that we believe could be associated with B. Further, we observe an elongated filamentary structure towards the east of knot F and South-East of the VLA sources, extending up to $\sim 0.5$ pc. We discuss these knots and structures further in Sect. 4.

Region 2 displays a complex morphology of $H_2$ arcs, reminiscent of terminal bow-shock heads observed towards multiple jets, such as

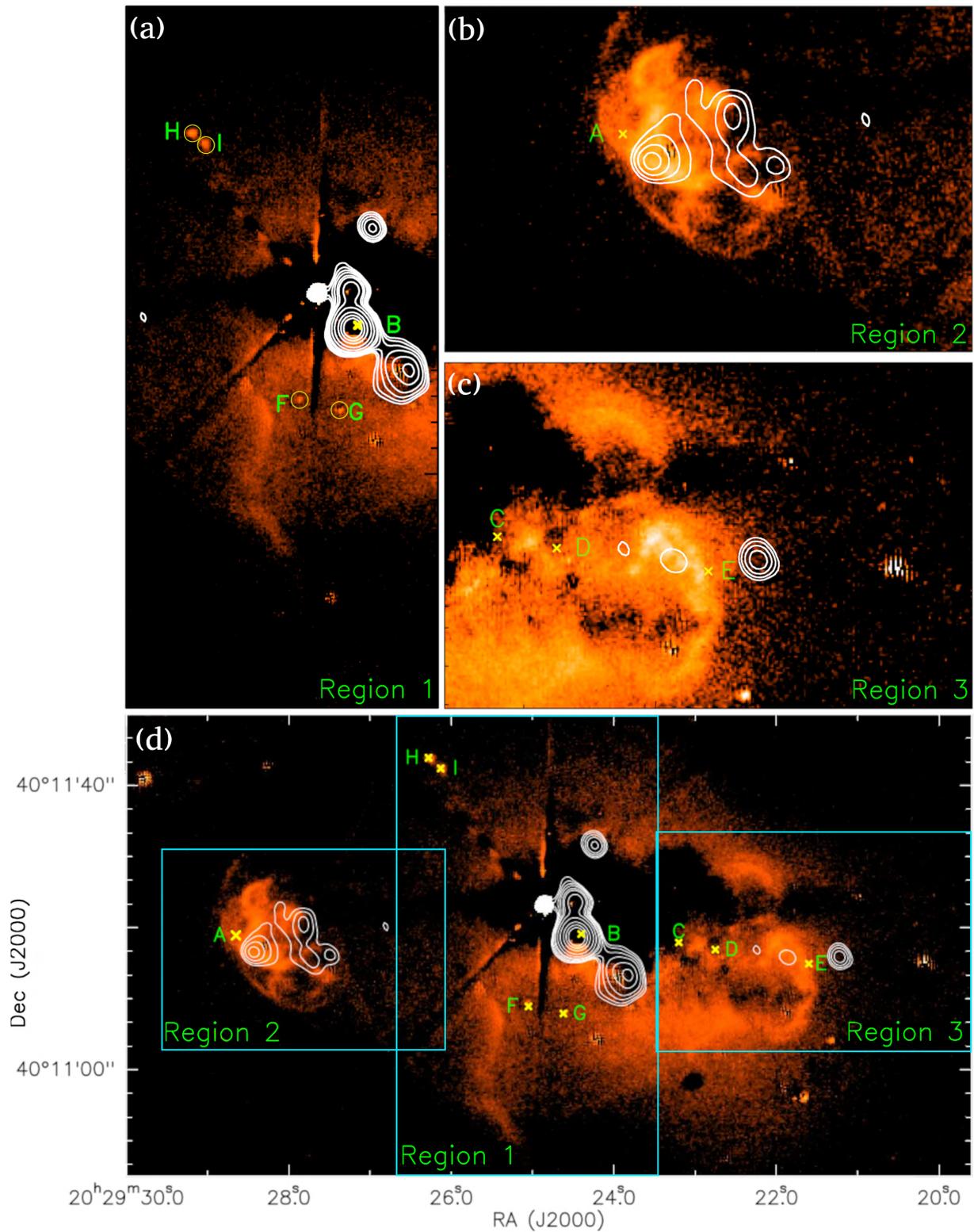

**Figure 4.** The UKIRT Widefield Infrared Survey for H$_2$ (UWISH2) image of the RAFGL2591 region overlaid with contours of 1280 MHz GMRT data. The yellow pointers locate approximate the peak positions of H$_2$ knots from Tamura & Yamashita (1992). Regions 1, 2 and 3 are enlarged for clarity.

HH 212, HH 34, HH 111, etc. (Zinnecker et al. 1998; Reipurth et al. 1986; Reipurth 1989). A faint and large-sized arc is seen towards the east of the multiple arcs, which likely represents the terminal head of bow shock from the jet. This is located at a distance of ∼0.75 pc from VLA 3. We discern the shocked regions as comprising of filaments, arcs and knot-like structures. The multiple arcs are reminiscent of bow shocks, with their heads pointing towards the eastern side, which is the direction of propagation of the jet. The brightest $H_2$ emission in Region 2 represents a bow-shock head, and this coincides with the emission of ionized gas, as observed at 1280 MHz, see Fig. 4 (b). This validates that GMRT-1 is associated with the jet emanating from the central driving source. We also perceive radio emission from the post-shocked gas behind the bow shock, some portions of which are devoid of $H_2$ emission. Towards Region 3, a filamentary arc is visible, pointing towards the west at a distance of ∼0.6 pc from VLA 3, similar to what was observed in Region 2 but in the opposite direction. We interpret this to be the terminal head of the western bow shock. Multiple knots are observed near the head of the bow shock. In this case, we observe that the radio emission appears ahead of the bow shock in the image. We also notice faint radio emission at 1280 MHz, at the $3\sigma$ level associated with the knots. The association of GMRT-2 with $H_2$ emission thus confirms its associated with the jet.

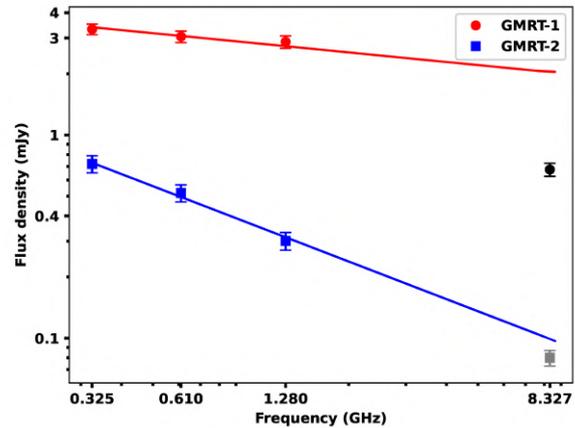

**Figure 5.** Spectral index plots of the jet lobes using flux densities at 325, 610, and 1280 MHz observed using GMRT. The red and black circles indicate flux densities of GMRT-1 obtained using GMRT and VLA at 8.4 GHz, respectively. The blue and grey squares indicate flux densities of GMRT-2 obtained using GMRT and VLA at 8.4 GHz, respectively.

### 3.3 Spectral indices of GMRT-1 and GMRT-2

We estimate the spectral indices of the jet lobes GMRT-1 and GMRT-2 by utilising the flux densities measured at the low-frequency bands of GMRT. The spectral indices help in ascertaining whether the emission is of thermal origin due to free-free emission, or non-thermal in nature due to synchrotron emission. This, in turn, can shed light on the properties of the region from where the radio emission is emanating. To determine the spectral indices, we first reconstruct the radio images having emission across similar spatial scales at the three frequencies. This is achieved by limiting the visibilities within a range of $0.6 - 36.5$ k$\lambda$, for the three frequency bands of GMRT. We estimate the flux densities associated with GMRT-1 and GMRT-2 at all three frequency bands, from apertures demarcated by the $5\sigma$ value in the 325 MHz image. The flux densities and spectral indices ($\alpha$) are reported in Table 3 and Fig. 5. The VLA 8.4 GHz flux densities of the lobes are also shown in Fig. 5. These have not been included in the estimation of spectral indices for the following reasons: (i) The absence of extended emission and other substructure(s) within the lobes in the image is the likely reason for low flux densities, and (ii) the uv coverage and range of the VLA observations are different from those of GMRT and hence, an adequate comparison of emission on similar spatial scales is difficult.

GMRT-1 has a relatively flat spectral index, $\alpha \sim -0.10 \pm 0.01$, while GMRT-2 displays a steeply negative value of spectral index $\alpha \sim -0.62 \pm 0.03$. The presence of synchrotron emission is evident from the spectral index value of GMRT-2. On the other hand, the relatively flat spectral index of GMRT-1 is plausibly explicable as thermal emission from shock-heated gas excited by the jet. We note that these represent a sort of average over the regions defined by the apertures that correspond to sizes 0.3 pc and 0.1 pc for GMRT-1 and GMRT-2, respectively. While a better approach would be to construct the spectral index maps towards these regions, the relatively low signal-to-noise ratio prohibits the accurate determination of spectral indices locally across regions.

### 3.4 Modelling radio emission from GMRT-1 and GMRT-2

In order to obtain estimates of the physical properties in regions of ionised gas emission in the radio lobes GMRT-1 and GMRT-2, we employ the numerical model of Mohan et al. (2022) and replicate the observed low-frequency radio spectra of GMRT-1 and GMRT-2. The model incorporates a generalised jet/lobe geometry of thermal emission and assimilates both thermal free-free and non-thermal synchrotron emission at the edges of the jet/lobe and terminal regions where the jet/lobe interacts with the ambient medium. The schematic of the model is shown in Fig. 6. The model provides improvements over the analytical jet model of Reynolds (1986) by taking into consideration larger opening angles, intermediate optical depth, lateral variation of ionization fraction, and feasibility of including synchrotron emission. The model can be utilised for radio emission from both, thermal jets close to the exciting sources, and lobes detected farther away. We apply it to the radio lobes in this case and estimate the total radio emission as a function of frequency. More details of the model, as well as the radiative transfer calculations, can be found in the paper by Mohan et al. (2022).

The model has a number of parameters: (i) $n_0$ represents the density of the lobe at the edge located closest to the exciting source - this distance is $r_0$, (ii) $\theta$ is the half-opening angle of the lobe and $\delta\theta$ represents the angular thickness of the shocked region with combined thermal and non-thermal emission, (iii) $q_n$, $q_x$ and $q'_x$ represent the power law indices of the radial variation of number density, radial variation of ionization fraction, and lateral variation of ionization fraction, respectively, (iv) $p$, $B$ and $\eta_{rel}$ represent parameters of non-thermal emission in the terminal regions: power-law index of electron population, the magnetic field and the fraction of electrons that are relativistic and producing synchrotron emission, respectively. We assume that the lobes are fully ionized at $r_0 = 0.6$ pc, are at a temperature of $10^4$ K, and are conical in shape, i.e. with a constant opening angle.

Amongst the above parameters, few are incorporated from previous observations towards RAFGL2591 while the remaining parameters are investigated in a range characteristic to jets by using a suitable grid as listed in Table 4. The chi-square minimization method is used to select the parameter set that best fits the data. We note that our sole objective in employing this model is to obtain order-of-magnitude estimates of parameters towards the lobes. The distance of the jet is taken as 3.33 kpc (Rygl et al. 2012), and the inclination angle of jet is considered as 20° with respect to the plane of the sky (Jiménez-Serra et al. 2012). We have considered the magnetic field to be 0.8 mG for GMRT-2 as determined in Sect. 4.2, while for GMRT-1, we have assumed a nominal value of 0.5 mG. The sizes of the radio lobes and opening angles are estimated by deconvolving the radio beam from the lobes. The lengths of the conical lobes are 0.32 and 0.16 pc for GMRT-1 and GMRT-2, respectively, while the opening angles are 12° and 4°, respectively. We impress upon the fact that the opening angles in the case of radio lobes far away from the jet are used solely to describe the regions of ionised gas in the lobes and may not represent the true opening angle of the jet. Based on these values, the parameters corresponding to the least $\chi^2$ values are given in Table 5, and the best-fit model radio spectra are shown in Fig. 7. We discuss each individual lobe below.

**Table 3.** Details of the radio lobes GMRT-1 and GMRT-2 including flux densities and spectral indices.

| Source | $\alpha_{J2000}$ (hms) | $\delta_{J2000}$ (° ′ ″) | $\theta_{src}$ (″×″) | 325 MHz (mJy) | 610 MHz (mJy) | 1280 MHz (mJy) | Spectral Index ($\alpha$) |
|---|---|---|---|---|---|---|---|
| GMRT-1 | 20:29:27.95 | +40:11:14.1 | 20 × 18 | 3.3 ± 0.2 | 3.1 ± 0.2 | 2.8 ± 0.2 | -0.10 ± 0.01 |
| GMRT-2 | 20:29:21.80 | +40:11:12.6 | 10 × 8 | 0.72 ± 0.07 | 0.51 ± 0.05 | 0.30 ± 0.03 | -0.62 ± 0.03 |

**Table 4.** The grid of input parameters used for radio lobe model calculations.

| Parameter | Description | Values for GMRT-1 | Values for GMRT-2 |
|---|---|---|---|
| $n_0$ | Number density (cm$^{-3}$) | 100, 500, 1000, 1500 | 100, 500, 1000, 1500 |
| $q_n$ | Power-law index of radial number density variation | -1, -1.5, -2, -2.5 | -1, -1.5, -2, -2.5 |
| $q_x$ | Power-law index of radial ionization fraction variation | -1, -2, -3, -4, -5 | -4, -5 |
| $q'_x$ | Power-law index of lateral ionization fraction variation | -1, -2, -3, -4, -5 | -2, -3, -4, -5 |
| $\eta_e^{\rm rel}$ | Fraction of relativistic electrons | 0, 0.1, 0.01 | $10^{-4}, 10^{-5}, 10^{-6}, 10^{-7}$ |
| $p$ | Power-law index of non-thermal electron population | 2.2, 2.3, 2.4, 2.5, 2.6 | 2.2, 2.3, 2.4, 2.5, 2.6 |
| $\delta\theta$ | Angular thickness of shocked region (°) | 0.01, 0.05, 0.1 | 0.01, 0.05, 0.1 |

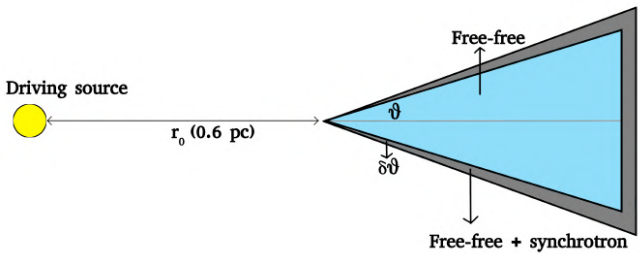

**Figure 6.** Schematic diagram of the model radio lobe with constant opening angle. The blue region of the lobe represents the inner (near the jet-axis) fully thermal jet. $r_0$ represents the distance of the lobe from the central source. The grey region of the lobe represents the geometrically thin region with shocked material that includes contribution from both, thermal and non-thermal emission.

### 3.4.1 GMRT-1

The GMRT-1 jet lobe is located at a projected distance of 0.6 pc from the central driving source. The best-fit model spectrum for this lobe, shown in Fig. 7(a), fits the observational data fairly well. The peak of the model spectrum has a flux density of 3.2 mJy at a turnover frequency of 310 MHz. The spectral index of the model observed between 325 and 1280 MHz is -0.10 and 0.98 between 10 and 100 MHz. The best-fit value for the number density is 1000 cm$^{-3}$. There are no previous estimates of the number densities towards the eastern region; hence a direct comparison is difficult. However, the number density obtained matches the observations of other jet lobes farther from exciting sources as observed in HH 34 and HH 80-81 (Buehrke et al. 1988; Heathcote et al. 1998). The variation of the number density towards the radial direction is given by the power-law index of the number density profile. This has been quantified as $q_n \sim -2$ using the obtained model spectrum. Previous observational studies and models have predicted a density profile with $q_n$ in the range of -1 to -2 for the circumstellar material observed around YSOs (van der Tak et al. 1999; Hogerheijde et al. 1999). The radial number density profile ($q_n \sim -2$) for the lobe is reasonable considering the fact that this lobe is located farther away (~ 0.6 pc) from the driving source, where the ambient density will be significantly lower than the immediate neighbourhood of the central YSO. In addition, it is also observed that the corresponding lobe of the jet which, has moved out of the molecular cloud into a region of lower density. Given this, we believe that the $q_n$ value obtained for this lobe is feasible. Due to internal shocks brought on by changes in flow velocity, the jet material is ionised mainly in the radial direction. A large value of $q_x \sim -2$ suggests that ionisation is substantially lower at larger radial distances within the lobe. $q'_x$ represents the level of ionisation brought on by shocks on the lateral margins of the jet when it collides with the surrounding medium. Likewise, a large value of $q'_x \sim -3$ points towards steeply diminishing ionisation in the lateral direction.

The model parameters $\eta_e^{\rm rel}$ and $p$, represent the effectiveness of shock in producing a relativistic electron population. As the best-fit model gives the relativistic electron population to be zero, the parameters $\delta\theta$ and $p$ are redundant. Thus, GMRT-1 is modelled as a radio lobe whose observed spectral index is explicable on the basis of thermal free-free emission only.

**Table 5.** Best-fit radio lobe model parameters for GMRT-1 and GMRT-2.

| Parameter | Description | GMRT-1 | GMRT-2 |
|---|---|---|---|
| $n_0$ | Number density (cm$^{-3}$) | 1000 | 100 |
| $q_n$ | Radial number density variation | -2 | -2 |
| $q_x$ | Radial ionization fraction variation | -2 | -3 |
| $q'_x$ | Lateral ionization fraction variation | -3 | -2 |
| $\eta_e^{\rm rel}$ | Fraction of relativistic electrons | 0 | $10^{-6}$ |
| $p$ | Non-thermal electron population | - | 2.3 |
| $\delta\theta$ | Angular thickness of shocked region (°) | - | 0.1 |

### 3.4.2 GMRT-2

The lobe GMRT-2 is located ∼ 0.6 pc from the primary driving source. The observed GMRT flux densities are reasonably well-fitted by the best-fit model spectra for this lobe, which is displayed in Fig. 7(b). The peak of the model spectrum has a flux density of 0.9 mJy at the turnover frequency of 126 MHz. The model's observed spectral indices are -0.62 between 325 and 1280 MHz, and 2.00 between 10 and 50 MHz.

For the number density, the best-fit value obtained is 100 cm$^{-3}$. The number densities towards the western side of the jet/outflow have been estimated before by Poetzel et al. (1992) with values ranging between 250 to 2000 cm$^{-3}$ across distances in the range 0.3 – 0.5 pc. The lower number density obtained from the model corroborates with the plausibly lower column density of the molecular clump that can be inferred from the Herschel 250 $\mu$m image (see Sect. 4.3 for more details). The best-fit model gives $q_n \sim 2$ for GMRT-2, similar to the case of GMRT-1. Inferring that the strength of internal shocks that might ionise the jet material is waning, large values of $q_x \sim -3$ show that the degree of ionisation is significantly lower across the lobe, suggesting the compact nature of emission. The relative contribution of synchrotron emission to the overall radio spectrum is significantly influenced by the model parameters $\eta_e^{\rm rel}$ and $p$. The shocked region in GMRT-2 has an angular thickness of 0.1° with a relativistic electron population of $\eta_{rel} \sim 10^{-6}$ and a power law index of $p \sim 2.3$ for non-thermal electron population. The relativistic electron population of $10^{-6}$ is comparable to the values obtained from the results of simulation studies in the context of diffusive shock acceleration (Berezhko & Ellison 1999; Padovani et al. 2016). The parameter $p \leq 2.4$ could imply that the majority of the particle kinetic energy is carried by those particles with energies significantly higher than the average. The obtained value for $p \sim 2.3$ is typical of shocks that are capable of generating non-thermal electron population in YSO jets (Araudo et al. 2021). Thus the observed flux densities of GMRT-2 are well-modelled by radio emission, having a dominant contribution from non-thermal synchrotron processes.

## 4 DISCUSSION

From the GMRT radio maps, we discern that both the radio lobes are equidistant (∼ 0.6 pc) from VLA 3, which suggests both the jet lobes must be driven by a single source. These radio lobes lie at position angles (PA) of 100° and 260° with respect to VLA 3, respectively. The latter is in accordance with the PA of 260° from H$_2$ knots seen towards the western side (Poetzel et al. 1992) and 270° of the blue CO outflow lobe (Hasegawa & Mitchell 1995). The radio lobe extension towards the eastern side of VLA 3 at 3.6 cm by Johnston et al. (2013) has a PA of 100°, which is consistent with the PA of GMRT-1. Multiple tracers have identified VLA 3 to the driving source of the large-scale jet / outflow (Poetzel et al. 1992; Hasegawa & Mitchell 1995; Gieser et al. 2019). The jet lobes are not diametrically opposite and show a mirror-symmetry, which we discuss further in Sect. 4.4.

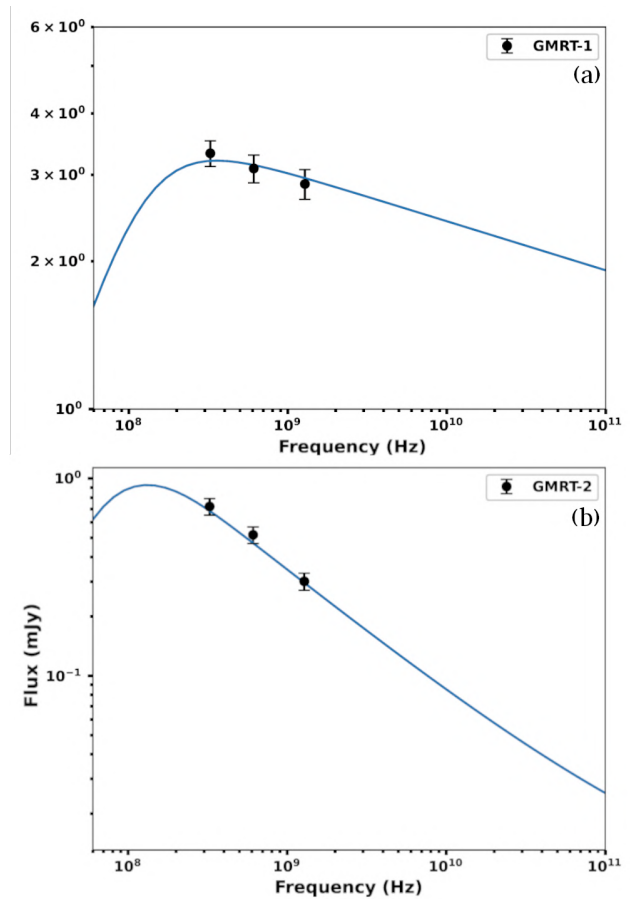

**Figure 7.** The flux densities of (a) GMRT-1 and (b) GMRT-2. The circles represent the GMRT flux densities, while the solid lines denote the best-fit model in each case.

We have been able to detect all the 5 VLA sources in our low-frequency radio images. However, it is not possible to disentangle their individual flux densities at all frequencies due to the effects of resolution. VLA 1, 2 and 3 are not resolved at 610 and 325 MHz, while VLA 4 and 5 are resolved at these frequencies. The integrated flux densities of these sources are higher at 1280 MHz as compared to 610 MHz. The 610-1280 MHz spectral indices of VLA 4 and VLA 5 are +0.9 and +0.4, respectively, suggesting the thermal nature of these regions. This is in accordance with the observations of Trinidad et al. (2003) that VLA 1, 2, 4 and 5 are HII regions.

### 4.1 GMRT-1

The flat spectral index of the eastern lobe GMRT-1 indicates the presence of thermal free–free emission as a result of shock-ionisation. As already seen, this radio lobe is associated with multiple H$_2$ emission arcs in the vicinity, and we note that the cometary radio lobe overlaps with the most intense H$_2$ emission knots in this region, with the head pointing towards the east. The terminal bow shock appears to lie ahead of the radio lobe by ∼ 1″ in the image. This indicates that the radio emission is likely to be produced in the ionised gas in the post-shock region. However, there exists the possibility that the ionised gas emission lies in the jet head and its appearance as behind the bow-shock head is the effect of projection as the red-shifted jet lobe in the east is pointed away from us, by an inclination angle of $i \sim 20°$ (Jiménez-Serra et al. 2012).

The bow shock head, as observed in $H_2$, extends for about 25″ in length and comprises inner knots and filamentary structures. This eastern lobe appears to be on the verge of breaking out of the molecular cloud of RAFGL2591. This is evident from the comparison of $H_2$ emission with cold dust emission associated with the molecular cloud, probed with Herschel 250 $\mu$m emission. Figure 8 displays the location of the radio emission GMRT-1 vis-a-vis the cold dust emission. As the jet propagates into the inhomogeneous medium of the molecular cloud, there is a tendency for gas to get compressed into a thin shell near the bow-shock and fragment due to instabilities that develop here (Blondin et al. 1990; Stone & Norman 1994).

The shape and width of the bow shock depend on the relative density of the jet with respect to the ambient medium, with a lower density ambient medium giving rise to a wider flow at the jet-head (de Gouveia dal Pino & Birkinshaw 1996). The velocity of the bow shock can be estimated using the equation, $v_{bs} \approx v_j \left[1 + (\eta\alpha)^{-1/2}\right]^{-1}$ (de Gouveia dal Pino & Benz 1993), where $v_j$ is the jet velocity and $v_{bs}$ is the velocity of the advancing bow shock. $\eta = \frac{n_j}{n_{mc}}$ represents the ratio of jet number density to the density of the ambient molecular cloud, while $\alpha$ gives the square of the ratio of the radius of the jet beam to the radius of the jet head bow-shock. For the case of bow-shock associated with GMRT-1, we take $v_j \sim 500$ km/s (Poetzel et al. 1992). We take the density of jet $n_j \sim 1000$ cm$^{-3}$ from our model, and consider the molecular cloud density range $n_{mc} \sim 250 - 2000$ cm$^{-3}$ based on western lobes (Poetzel et al. 1992), which gives $\eta \sim 2.5 - 20$. The radius of the jet head bow-shock as measured from $H_2$ emission map is $\sim 13.5″$, and the radius of the jet beam from the size of radio lobe GMRT-1 is 9.5″, which gives $\alpha \sim 2$. With these values, the bow shock velocity would lie in the range $\sim 120 - 230$ km s$^{-1}$.

### 4.2 GMRT-2

The western lobe GMRT-2 exhibits non-thermal characteristics with a steep spectral index of $-0.62$. The $H_2$ emission comprises of nebulous emission with a series of knots across the full extent of the western side of the jet. In addition to being compact, the GMRT-2 lies $\sim 3″$ ahead of the terminal head of the associated bow-shock seen in $H_2$, unlike the case of GMRT-1. As in the East, we notice an arc-shaped structure in $H_2$ towards the west as well. The extended $H_2$ emission is explicable on the basis of the fact that the western side of the jet is blue-shifted, and therefore the column density along the line-of-sight is likely to be lower. This would also explain the loops discerned in the optical and infrared towards this side (Tamura & Yamashita 1992).

The velocity of shocked $H_2$ gas towards the western lobe has been estimated to be $\sim 500$ km s$^{-1}$ (Poetzel et al. 1992). Taking this to be the velocity of the jet, we estimate a jet dynamical timescale of 1180 yr, considering the position of the radio lobe as the outermost extent of the jet. Unlike the case of GMRT-1, GMRT-2 appears to be impinging on a higher-density molecular gas clump. This can be appreciated by comparing the location of the lobe with the distribution of cold dust emission at 250 $\mu$m, shown in Fig. 8. We note that the location of the radio lobe is such that it appears to be moving towards a higher density clump located towards the South-West of the molecular cloud. This would probably explain the formation of a strong shock that ionises the gas. The presence of a radio lobe ahead of the $H_2$ emission indicates that $H_2$ emission arises from the cooler wing of the bow shock, possibly in the post-shock cooling zone.

The presence of non-thermal synchrotron emission in radio lobes of jets has been attributed to the acceleration of electrons to relativistic velocities in the presence of magnetic field by the process of diffusive shock acceleration across the shock front (Masqué et al. 2019; Rodríguez-Kamenetzky et al. 2019). We estimate the strength of the magnetic field under the assumption of minimization of the total energy content of the synchrotron source. This approximately corresponds to the equipartition of energy between the magnetic field and the relativistic particles. We use the following equation to estimate the magnetic field towards this lobe (Miley 1980).

$$\left(\frac{B_{eq}}{\text{gauss}}\right) = 5.69 \times 10^{-5} \left[\frac{1+K}{\eta(\sin\phi)^{3/2}(\alpha+1/2)} \left(\frac{\text{arcsec}^2}{\theta_x\theta_y}\right)\right]^{2/7} \times \left[\left(\frac{\text{kpc}}{s}\right) \left(\frac{F_0}{\text{Jy}}\right) \frac{\left(v_2^{\alpha+1/2} - v_1^{\alpha+1/2}\right)}{v_0^\alpha(\alpha+1/2)}\right]^{2/7} \quad (1)$$

Here $\chi$ is the ratio of the energy of the heavy particles to that of electrons observed in cosmic rays near the earth, and is taken as 40 (Simpson et al. 1983). The angle between the line-of-sight and the magnetic field is $\phi$, and we take $\sin\phi \sim 0.5$ (Johnston et al. 2013; Masqué et al. 2019). The path length through the source in the line-of-sight is represented as $s$. $F_0$ is the radio flux density at frequency $v_0$, which is taken as 325 MHz; $v_1$, and $v_2$ are the cutoff frequencies, which are taken as 0.1 GHz and 100 GHz, while $f$ is the filling factor of the source (assumed to be 0.5). With these values, we estimate a value of $B_{eq} \sim 0.8$ mG for GMRT-2. A comparison with magnetic field estimates derived from radio condensations towards other jets indicates that our value is consistent with values found for other lobes, in the range $0.1 - 1$ mG (Kuiper et al. 2010; Ainsworth et al. 2014; Masqué et al. 2015; Vig et al. 2018; Masqué et al. 2019).

In order to understand the origin of the synchrotron emission, we estimate the cooling time for relativistic electrons in GMRT-2 through this process. We use the following expression to estimate the cooling time $t_c$ (Longair 2011).

$$t_c(\text{yr}) \sim 2.3 \times 10^4 v^{-1/2} B^{-3/2} \quad (2)$$

Here, $v$ is the frequency of observations in Hz, and $B$ is the magnetic field in Gauss, which we take as 0.8 mG. Using $v = 325$ MHz, the cooling time for electrons in GMRT-2 is $5 \times 10^4$ yrs. When compared with the jet dynamical timescale of 1180 yrs, we find that the electrons would not have had sufficient time to cool, implying the adiabatic nature of the shock-front, leading to relativistic acceleration of the electrons.

### 4.3 Outflows from RAFGL2591

Two outflows have been observed from the vicinity of VLA 3 and we discuss them in the context of our results in this section.

#### 4.3.1 East-West Outflow

The prominent jet and outflow are in the East-West direction, which has been probed by numerous studies (Lada et al. 1984; Poetzel et al. 1992; Tamura & Yamashita 1992; Hasegawa & Mitchell 1995; Johnston et al. 2013; Gieser et al. 2019). In this work, we report the detection of radio lobes and extended $H_2$ emission towards both the East and West directions. On close inspection towards VLA 3, we note emission extending along PA ∼ 252° from the west of VLA 3 in the UKDISS J and H-band images. This is also observed in the speckle interferometric image of VLA 3 by Preibisch et al. (2003), and we interpret this near-infrared continuum emission as arising from the hot dust in the jet cavity walls. The sizes of GMRT-1 and GMRT-2 suggest opening angles of the lobes to be 24° and 8°, respectively. Although the dust loop cavities suggest larger opening angles, it is clear that the driver of the outflow is a jet rather than a wide-angled stellar wind.

We obtain estimates of the mass-loss rate by utilising the total mass of the gas in the lobes and the dynamical time scale. We consider GMRT-1 for this purpose. The best-fit model number densities (see Sect. 3.4) and its variation across the lobe are exploited to obtain the total mass in the lobe of GMRT-1, which is 0.58 $M_\odot$. Considering a dynamical time scale of 1180 yrs, we estimate the mass loss rate to be $5 \times 10^{-4}$ $M_\odot$ yr$^{-1}$. This is consistent with the mass-loss rate of the associated CO outflow, $7.0 \times 10^{-4}$ $M_\odot$ yr$^{-1}$ (Hasegawa & Mitchell 1995). An alternate measure of the ionised mass-loss rate from the jet is given by Johnston et al. (2013) in the proximity of VLA 3, and they obtain a value in the range $0.7 - 1.0 \times 10^{-5}$ $M_\odot$ yr$^{-1}$. Our value is an order of magnitude higher than this measurement and represents the mass-loss rate associated with total gas content. The value of mass-loss rate from RAFGL2591 is towards the higher end of mass-outflow rates observed towards other massive protostellar objects: $10^{-5} - 10^{-6}$ $M_\odot$ yr$^{-1}$ (Carrasco-González et al. 2012; Guzmán et al. 2012). This is about 3-4 orders of magnitude larger than those observed towards low-mass protostars (Podio et al. 2006).

*4.3.2 NE-SW Outflow*

A close-up view of VLA 3 by Gieser et al. (2019) has revealed the presence of an alternate outflow in the region with the blue and red-shifted lobes towards North-East and South-West directions through the tracers SiO and SO, and we designate this as NE-SW outflow. This is also the direction in which we have detected the H$_2$ knots H, and I oriented linearly with PA∼ 42° with respect to VLA 3, at a distance of ∼ 0.4 pc. We believe that these H$_2$ knots represent the shocked gas emission associated with the NE-SW jet, which is at PA∼ 45° in the jet tracer SiO.

*4.3.3 Drivers of these outflows*

The presence of two jets (E-W and NE-SW) from VLA 3 implies the presence of two protostars lying in close vicinity. We note that the geometric mean of the red and blue lobes of the SiO jet as well as SO outflow is towards the North-East of VLA 3. We, therefore, postulate that a binary companion of VLA 3 that lies towards the North-East (< 2″ based on molecular tracers), is giving rise to the NE-SW jet. We speculate that AFGL2591-2 lying 1.9″ to NE (Beuther et al. 2018) is a possible driving source of the NE-SW jet. An alternate possibility is the fragment/core A located ∼ 0.2 pc to the North-East of VLA 3 (Suri et al. 2021). In either case, we anticipate that the core driving the NE-SW jet must have its own individual disk, and therefore the core must be non-coplanar with the disk of VLA 3 driving the E-W jet, leading to the two distinct jets with a projected angular separation of ∼ 50°.

A schematic of the scenario with both outflows is shown in Fig. 9. Higher resolution observations of the jet and detailed investigation into the likely exciting sources can help identify the origin of the NE-SW jet.

### 4.4 Misalignment of East-West jet

The sizes of GMRT-1 and GMRT-2 are relatively compact in comparison to the extent of associated H$_2$ emission, which can help us in gauging the direction of jet propagation accurately. As mentioned in Sect. 4.3, the equidistant lobes of the E-W jet are at PA angles of 100° and 260°. The PA of GMRT-1 is identical to that estimated by Johnston et al. (2013) by the ionised radio jet closer to VLA 3, at about 3000 AU. On the western side, too, we note the extension of near-infrared continuum emission at separations as close as 3000 AU from VLA 3. A marginal extension of radio emission at 3.6 cm associated with VLA 3 is also observed towards the west in the same direction, and this can be seen from the inset in Fig. 8 (bottom).

This implies that the jet and counter-jet are misaligned with respect to each other by ∼ 20°, implying a Mirror or 'C' symmetry (Bally & Reipurth 2001; Bally et al. 2006). This is in contrast to the Point or 'S' symmetry also observed in jets (Margon et al. 1979; Monceau-Baroux et al. 2015) where the bending of the jet is symmetric but in opposite directions. Fendt & Zinnecker (1998) propose various scenarios of jet-bending and conclude that the presence of binary/multiple systems and Lorentz forces on the magnetic field of the jet as likely causes. The binary companion premise has been analysed by a number of researchers to explain the observed symmetries via the binary orbital motion, precession of disk due to tidal interaction with a binary companion, and the effects of co-planar and non-coplanar binary companion relative to the disk of the jet source (Shepherd et al. 2000; Crocker et al. 2002; Beltrán et al. 2016; Dominik et al. 2021). We note that VLA 3 is part of a cluster of young stellar objects and cores (VLA 1, VLA 2, VLA 4, VLA 5, AFGL2591-2, and AFGL2591-3) with additional sub-solar fargments in the vicinity. The presence of these cores/fragments is a likely cause of jet misalignment.

An alternate possibility that has been proposed for the 'C' symmetry is the presence of deflecting effects that can veer away one of the lobes of the jet leading to its misalignment with the counter jet (Canto & Raga 1995). According to Bally & Reipurth (2001), a neutral jet beam can be deflected by three different mechanisms: (i) a side wind acting directly on the jet, (ii) radiation pressure acting on dust entrained in the jet, and (iii) by the rocket effect operating on the illuminated side of a neutral jet beam. If the radiation field is anisotropic, the latter two forces may act via a jet cocoon to directly deflect the jet beam. In the case of the RAFGL2591 E-W jet, there is a possibility that one of the lobes is affected by supersonic side winds or other deflecting effects, leading to its relative misalignment. We note that most of the young stellar objects, such as those exciting the nearest HII regions VLA 1 and VLA 2, are powered by massive stars of single ZAMS spectral type B0.5 and B1, respectively (Johnston et al. 2013). It is possible that they could be the drivers of supersonic winds affecting the western jet. However, we note that they are located about 0.09 and 0.07 pc away, and it needs to be established if they could be responsible for the diverting effects that can bend the jet lobe at these distances.

Lastly, we consider the possibility of proper motion of the exciting source being responsible for the misalignment (Reiter et al. 2020). In this case, the driving source ejects bipolar knots of material which continue to travel ballistically while the driving source continues to move. If we consider this hypothesis as the cause of jet bending observed at 0.6 pc, then a deflection of 10° for each lobe over a dynamical timescale of ∼ 1180 yrs would suggest a proper motion of VLA 3 by 6.9″ during this time period. This proper motion would imply a velocity of 90 km/s or greater for VLA 3. This is highly

12    *Cheriyan et al.*

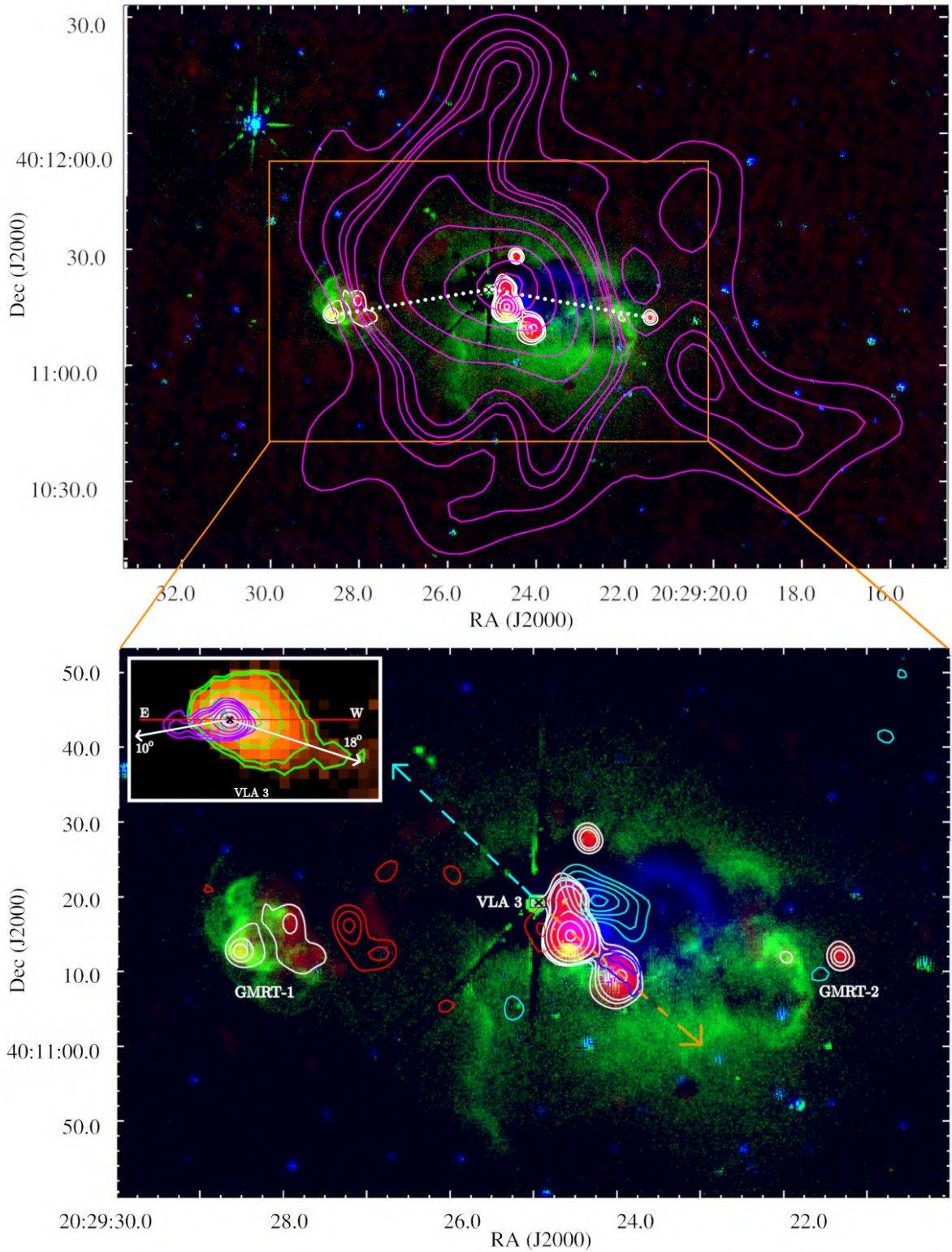

**Figure 8.** Three colour composite image of the RAFGL2591 region encompassing the E-W jet is shown in the top panel, with UKIDSS J band (blue), UWISH2 2.12 $\mu$m (green), and GMRT 1280 MHz radio image (red). The 1280 MHz contours are the same as shown in Fig. 2, while the magenta contours depict the 250 $\mu$m emission from Herschel. The white dotted lines (top panel) point to the radio lobes from VLA 3, highlighting the misalignment between them. The central region is enlarged in the bottom panel. Here, the CO outflow is shown as cyan and red contours, while the cyan and orange dashed arrows indicate the directions of the blue and red lobes of the NE-SW jet. The inset shown on the top left of this panel zooms close to VLA 3. The violet contours represent VLA 8.4 GHz emission overlaid over the UKDISS J band image, with the emission in the latter also highlighted through the green contours.

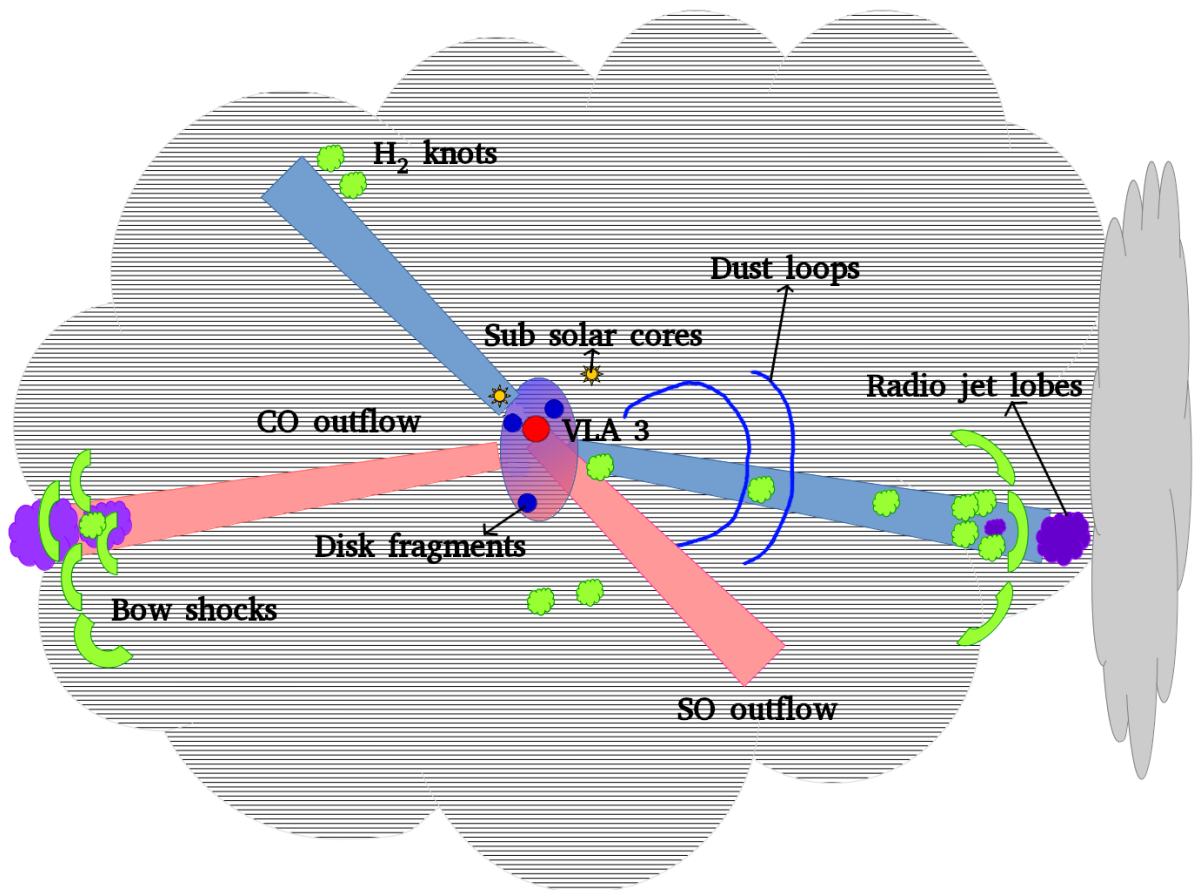

**Figure 9.** Schematic of the RAFGL2591 region showing various characteristics VLA 3 (shown as red source), not-to-scale. The yellow stellar sources indicate the cores identified by Beuther et al. (2018), and blue circles represent the subsolar fragments A, B, and C detected by Suri et al. (2021). The pale blue ellipse reveals the disk of the protostar. The two elongated pink and blue lobes represent the two jets/outflows associated with VLA 3. The dust loops in the region are shown as blue thin curved lines. The purple structures represent the radio lobes (GMRT-1 and GMRT-2), and green structures (arcs, knots) represent shocked $H_2$ emission. The cloud in which VLA 3 is embedded is shown with hatched lines, while the clump towards the western arm is denoted by solid grey, where the radio lobe GMRT-2 impinges. More details can be found in the text.

unlikely as it is embedded in the molecular cloud with no apparent cause that could lead to such high velocities.

Based on the above, we hypothesize that binary companion(s) and/or supersonic sidewinds as feasible causes for the misalignment of the E-W jet lobes, and rule out proper motion of VLA 3 as a likely origin.

## 5 CONCLUSIONS

The conclusions from the current work are summarised in the following points.

(i) We have carried out a low radio frequency mapping of the RAFGL2591 star-forming region across three frequency bands: Band 3 (250-500 MHz), Band 4 ( 550-850 MHz) and Band 5 (1050-1450 MHz) using the Giant Metrewave Radio Telescope (GMRT). For the first time, we detect radio lobes of the E-W jet in all bands at a distance of 0.6 pc and designate them as GMRT-1 and GMRT-2. The spectral indices of GMRT-1 and GMRT-2 are $\alpha \sim -0.10$ and $\sim -0.62$, suggesting thermal and synchrotron emission, respectively, from these lobes. Emission from the five VLA sources towards the central region are also detected at the lower frequencies.

(ii) The UWISH2 $H_2$ image displays several arcs, filamentary structures and knots in regions away from the exciting source(s). Multiple arcs and nebulous emission are observed towards GMRT-1 and GMRT-2, suggesting interaction of the jet with the ambient medium. In addition to previously detected $H_2$ knots, for the first time, we have detected three additional knots in the vicinity of the VLA sources.

(iii) Using the equipartition of energy, the magnetic field estimated using the radio flux densities at 325 MHz is ~0.8 mG for GMRT-2 with a cooling time of $5 \times 10^4$ yrs.

(iv) Considering the mass of the jet lobe, GMRT-1 and a dynamical time scale of $\sim$ 1180 yrs, the mass loss rate has been estimated as $5 \times 10^{-4}$ $M_\odot$ yr$^{-1}$.

(v) The prominent jet is the one in the East-West direction consistent with $C^{18}O$ outflow lobes, GMRT-1 and GMRT-2, as well as $H_2$ structures towards the lobes. An alternate outflow observed earlier from VLA 3 in the North-East and South-West directions is consistent with the alignment of two new $H_2$ knots reported here. Both these outflows are believed to be excited from region close to VLA 3, and we speculate about the presence of another protostar (AFGL2591-2 or fragment/core A) in the close neighbourhood of VLA 3.

(vi) The East-West jet exhibits a misalignment having a reflection symmetry, with a relative bending of $\sim$ 20° between the lobes. This

could be due to the combined effect of the precession caused by a binary partner and a supersonic sidewind from one of the nearby VLA exciting sources.

(vii) We have applied a model of radio emission lobes associated with a jet to GMRT-1 and GMRT-2 to constrain the ranges of physical parameters in these lobes. We find the number densities of GMRT-1 and GMRT-2 to be 1000 and 100 cm$^{-3}$ with a power law index for radial number density variation, $q_n \sim 2$ for both the lobes. The best-fit model of emission from GMRT-1 includes only thermal emission, while GMRT-2 has been modelled as having a dominant contribution from non-thermal synchrotron emission.

(viii) This work shows the larger extent of E-W radio jet from RAFGL2591 using low radio frequencies. We, therefore, emphasize on the necessity of observing a large sample of protostellar jets at low radio frequencies.

**ACKNOWLEDGEMENTS**

We thank the anonymous referee for providing suggestions that have improved the clarity of the manuscript. We thank the staff of GMRT, who made the radio observations possible. GMRT is run by the National Centre for Radio Astrophysics of the Tata Institute of Fundamental Research. We also thank UKIRT for the UKIDSS-WFCAM and UWISH2 images. The UKIRT is supported by NASA and operated under an agreement among the University of Hawaii, the University of Arizona and Lockheed Martin Advanced Technology Center; operations are enabled through the cooperation of the Joint Astronomy Centre of the Science and Technology Facilities Council of the UK. We also made use of data products from Herschel (ESA space observatory). We also thank Katharine G. Johnston for providing the 8.4 GHz VLA maps and KARMA CO data cubes.

**DATA AVAILABILITY**

The data underlying this paper will be shared on reasonable request to the corresponding author.

**REFERENCES**

Ainsworth R. E., Scaife A. M. M., Ray T. P., Taylor A. M., Green D. A., Buckle J. V., 2014, ApJ, 792, L18
Anglada G., 1995, in Lizano S., Torrelles J. M., eds, Revista Mexicana de Astronomia y Astrofisica Conference Series Vol. 1, Revista Mexicana de Astronomia y Astrofisica Conference Series. p. 67
Anglada G., Rodríguez L. F., Carrasco-González C., 2018, A&ARv, 26, 3
Araudo A. T., Padovani M., Marcowith A., 2021, MNRAS, 504, 2405
Arce H. G., Sargent A. I., 2004, ApJ, 612, 342
Arce H. G., Shepherd D., Gueth F., Lee C. F., Bachiller R., Rosen A., Beuther H., 2007, in Reipurth B., Jewitt D., Keil K., eds, Protostars and Planets V. p. 245 (arXiv:astro-ph/0603071), doi:10.48550/arXiv.astro-ph/0603071
Bally J., 2016, ARA&A, 54, 491
Bally J., Lane A. P., 1991, in Elston R., ed., Astronomical Society of the Pacific Conference Series Vol. 14, Astronomical Society of the Pacific Conference Series. pp 273–278
Bally J., Reipurth B., 2001, ApJ, 546, 299
Bally J., Licht D., Smith N., Walawender J., 2006, AJ, 131, 473
Beltrán M. T., Cesaroni R., Moscadelli L., Sánchez-Monge Á., Hirota T., Kumar M. S. N., 2016, A&A, 593, A49
Berezhko E. G., Ellison D. C., 1999, ApJ, 526, 385
Beuther H., Shepherd D., 2005, in Kumar M. S. N., Tafalla M., Caselli P., eds, Astrophysics and Space Science Library Vol. 324, Astrophysics and Space Science Library. p. 105 (arXiv:astro-ph/0502214), doi:10.1007/0-387-26357-8_8
Beuther H., et al., 2018, A&A, 617, A100
Blondin J. M., Fryxell B. A., Konigl A., 1990, ApJ, 360, 370
Buehrke T., Mundt R., Ray T. P., 1988, A&A, 200, 99
Campbell B., 1984, ApJ, 287, 334
Canto J., Raga A. C., 1995, MNRAS, 277, 1120
Carrasco-González C., Rodríguez L. F., Anglada G., Martí J., Torrelles J. M., Osorio M., 2010, Science, 330, 1209
Carrasco-González C., et al., 2012, ApJ, 752, L29
Cécere M., Velázquez P. F., Araudo A. T., De Colle F., Esquivel A., Carrasco-González C., Rodríguez L. F., 2016, ApJ, 816, 64
Crocker M. M., Davis R. J., Spencer R. E., Eyres S. P. S., Bode M. F., Skopal A., 2002, MNRAS, 335, 1100
Dominik R. M., Linhoff L., Elsässer D., Rhode W., 2021, MNRAS, 503, 5448
Fendt C., Zinnecker H., 1998, A&A, 334, 750
Froebrich D., et al., 2011, MNRAS, 413, 480
Gieser C., et al., 2019, A&A, 631, A142
Guzmán A. E., Garay G., Brooks K. J., Voronkov M. A., 2012, ApJ, 753, 51
Hasegawa T. I., Mitchell G. F., 1995, J. R. Astron. Soc. Canada, 89, 173
Heathcote S., Reipurth B., Raga A. C., 1998, AJ, 116, 1940
Hogerheijde M. R., van Dishoeck E. F., Salverda J. M., Blake G. A., 1999, ApJ, 513, 350
Jiménez-Serra I., Zhang Q., Viti S., Martín-Pintado J., de Wit W. J., 2012, ApJ, 753, 34
Johnston K. G., Shepherd D. S., Robitaille T. P., Wood K., 2013, A&A, 551, A43
Kuiper R., Klahr H., Beuther H., Henning T., 2010, ApJ, 722, 1556
Lada C. J., Thronson H. A. J., Smith H. A., Schwartz P. R., Glaccum W., 1984, ApJ, 286, 302
Lal D. V., Rao A. P., 2007, MNRAS, 374, 1085
Latif M. A., Schleicher D. R. G., 2016, A&A, 585, A151
Li Z.-Y., Nakamura F., 2006, ApJ, 640, L187
Livio M., 1997, in Wickramasinghe D. T., Bicknell G. V., Ferrario L., eds, Astronomical Society of the Pacific Conference Series Vol. 121, IAU Colloq. 163: Accretion Phenomena and Related Outflows. p. 845
Longair M. S., 2011, High Energy Astrophysics. .
Margon B., Ford H. C., Grandi S. A., Stone R. P. S., 1979, ApJ, 233, L63
Masqué J. M., Rodríguez L. F., Araudo A., Estalella R., Carrasco-González C., Anglada G., Girart J. M., Osorio M., 2015, ApJ, 814, 44
Masqué J. M., Jeyakumar S., Trinidad M. A., Rodríguez-Esnard T., Ishwara-Chandra C. H., 2019, MNRAS, 483, 1184
Meier D. L., Koide S., Uchida Y., 2001, Science, 291, 84
Miley G., 1980, ARA&A, 18, 165
Mohan S., Vig S., Mandal S., 2022, MNRAS, 514, 3709
Monceau-Baroux R., Porth O., Meliani Z., Keppens R., 2015, A&A, 574, A143
Nakamura F., Li Z.-Y., 2007, ApJ, 662, 395
Obonyo W. O., Lumsden S. L., Hoare M. G., Purser S. J. D., Kurtz S. E., Johnston K. G., 2019, MNRAS, 486, 3664
Olnon F. M., 1975, A&A, 39, 217
Padovani M., Marcowith A., Hennebelle P., Ferrière K., 2016, A&A, 590, A8
Podio L., Bacciotti F., Nisini B., Eislöffel J., Massi F., Giannini T., Ray T. P., 2006, A&A, 456, 189
Poetzel R., Mundt R., Ray T. P., 1992, A&A, 262, 229
Preibisch T., Balega Y. Y., Schertl D., Weigelt G., 2003, A&A, 412, 735
Ray T. P., Ferreira J., 2021, New Astron. Rev., 93, 101615
Reipurth B., 1989, Nature, 340, 42
Reipurth B., Bally J., Graham J. A., Lane A. P., Zealey W. J., 1986, A&A, 164, 51
Reiter M., Haworth T. J., Guzmán A. E., Klaassen P. D., McLeod A. F., Garay G., 2020, MNRAS, 497, 3351
Reynolds S. P., 1986, ApJ, 304, 713
Rodríguez-Kamenetzky A., Carrasco-González C., Araudo A., Torrelles J. M., Anglada G., Martí J., Rodríguez L. F., Valotto C., 2016, ApJ, 818, 27

Rodríguez-Kamenetzky A., Carrasco-González C., González-Martín O., Araudo A. T., Rodríguez L. F., Vig S., Hofner P., 2019, MNRAS, 482, 4687

Rygl K. L. J., et al., 2012, A&A, 539, A79

Sánchez-Monge Á., Kurtz S., Palau A., Estalella R., Shepherd D., Lizano S., Franco J., Garay G., 2013, ApJ, 766, 114

Sanna A., Reid M. J., Carrasco-González C., Menten K. M., Brunthaler A., Moscadelli L., Rygl K. L. J., 2012, ApJ, 745, 191

Schulz N. S., 2012, The Formation and Early Evolution of Stars. ., doi:10.1007/978-3-642-23926-7

Shang H., Liu C.-F., Krasnopolsky R., Wang L.-Y., 2023a, ApJ, 944, 230

Shang H., Krasnopolsky R., Liu C.-F., 2023b, ApJ, 945, L1

Shepherd D. S., Yu K. C., Bally J., Testi L., 2000, ApJ, 535, 833

Shimajiri Y., Takahashi S., Takakuwa S., Saito M., Kawabe R., 2008, ApJ, 683, 255

Shull J. M., Beckwith S., 1982, ARA&A, 20, 163

Simpson J. A., Wefel J. P., Zamow R., 1983, in International Cosmic Ray Conference. pp 322–325

Stone J. M., Norman M. L., 1994, ApJ, 420, 237

Suri S., et al., 2021, A&A, 655, A84

Swarup G., Ananthakrishnan S., Kapahi V. K., Rao A. P., Subrahmanya C. R., Kulkarni V. K., 1991, Current Science, 60, 95

Tamura M., Yamashita T., 1992, ApJ, 391, 710

Terquem C., Eislöffel J., Papaloizou J. C. B., Nelson R. P., 1999, ApJ, 512, L131

Trinidad M. A., et al., 2003, ApJ, 589, 386

Vig S., Veena V. S., Mandal S., Tej A., Ghosh S. K., 2018, MNRAS, 474, 3808

Zinnecker H., McCaughrean M. J., Rayner J. T., 1998, Nature, 394, 862

de Gouveia dal Pino E. M., Benz W., 1993, ApJ, 410, 686

de Gouveia dal Pino E. M., Birkinshaw M., 1996, ApJ, 471, 832

van der Tak F. F. S., van Dishoeck E. F., Evans Neal J. I., Bakker E. J., Blake G. A., 1999, ApJ, 522, 991

This paper has been typeset from a TeX/LaTeX file prepared by the author.